# Large Stark Tuning of InAs/InP Quantum Dots

*Shahriar Aghaeimeibodi[1], Chang-Min Lee[1], Mustafa Atabey Buyukkaya[1], Christopher J. K. Richardson[2], and Edo Waks[1,3,a]*

[1] Department of Electrical and Computer Engineering and Institute for Research in Electronics and Applied Physics, University of Maryland, College Park, Maryland 20742, United States

[2] Laboratory for Physical Sciences, University of Maryland, College Park, Maryland 20740, United States

[3] Joint Quantum Institute, University of Maryland and the National Institute of Standards and Technology, College Park, Maryland 20742, United States

ABSTRACT

InAs/InP quantum dots are excellent sources of telecom single-photon emission and are among the most promising candidates for scalable quantum photonic circuits. However, geometric differences in each quantum dot leads to slightly different emission wavelengths and hinders the possibility of generating multiple identical quantum emitters on the same chip. Stark tuning is an efficient technique to overcome this issue as it can control the emission energy of individual quantum dots through the quantum-confined Stark effect. Realizing this technique in InAs/InP quantum dots has previously been limited to shifts of less than 0.8 meV due to jumps in the emission energy because of additional charges at high electric field intensities. We demonstrate up to 5.1 meV of Stark tuning in the emission wavelength of InAs/InP quantum dots. To eliminate undesirable jumps to charged state, we use a thin oxide insulator to prevent carrier injection from the contacts, thereby significantly improves the tuning range of the Stark effect. Moreover, the single-photon nature and narrow linewidth of the quantum dot emission is preserved under a wide range of applied electric fields. Using photoluminescence intensity measurements and time-resolved lifetime spectroscopy we confirmed that this Stark tuning range is limited by carrier tunneling at high electric fields. This result is an important step toward integrating multiple identical quantum emitters at telecom wavelengths on-a-chip, which is crucial for realizing complex quantum photonic circuits for quantum information processing.

---

a) Corresponding Author, E-mail: edowaks@umd.edu

Epitaxial InAs quantum dots embedded in InP (InAs/InP) are excellent sources of telecom single photons.[1,2] Previous studies have demonstrated single-photon generation in these materials with large Purcell enhancement[3] and record high single-photon purity.[4] InAs/InP quantum dots can also act as indistinguishable single-photon sources at telecom wavelengths suitable for quantum communications.[5] Many applications in quantum information processing require multiple identical quantum emitters. However, the emission wavelength of individual quantum dots varies due to small differences in their shape and size, which makes it difficult to integrate multiple identical quantum emitters on the same chip. To overcome this spectral randomness, we need a method to tune the emission wavelength of individual dots over a broad spectral range without degrading the coherence properties of the emitter.

Stark tuning is an effective technique to tune the wavelengths of quantum dots. This technique uses an electric field to tune the emission wavelength using the quantum-confined Stark effect. In InAs/GaAs quantum dots emitting at wavelengths below 1 μm, Stark tuning has produced wavelength shifts of up to 25 meV with minimal degradation of emission properties,[6,7] enabling two different dots to show two-photon interference.[8,9] However, the extension of this approach to InAs/InP quantum dots that emit at telecom wavelength has been challenging. Previous Stark tuning demonstrations of InAs/InP quantum dots suffered from a limited tuning range of < 0.8 meV[10,11], because at high electric field intensities the dot gains additional charges that cause discrete jumps in the optical spectrum. Achieving large wavelength tuning ranges requires new device geometries that limit this charging effect.

In this letter, we demonstrate a large DC Stark shift in the emission wavelength of InAs/InP quantum dots at telecommunication wavelengths. We place the InAs quantum dots embedded in InP between two metallic plates that are separated by a $SiO_2$ buffer layer to avoid introduction of charges. An applied DC electric voltage shifts the emission of the quantum dot by more than 5 meV without any observed spectral jumps due to charging, providing a much larger tuning range than previously available for this material. Steady-state and time-resolved photoluminescence measurements show a clear tunneling effect introducing non-radiative decay processes that limit the emitter brightness at high voltages. Photon correlation measurements show robust anti-bunching behavior over a wide voltage range. Our results pave the way for developing large identical arrays of quantum emitters at telecom wavelengths. Such identical quantum emitters are highly desirable for quantum information applications, such as boson sampling,[12] quantum communication,[13] and optical quantum computing.[14–16] Furthermore, our technique is agnostic to the nature of the substrate, and could thus be incorporated in more complex hybrid photonic device structures that combine quantum dots with silicon photonics[17] or lithium niobate photonics.[18]

Figure 1a shows the proposed device, in which the InAs/InP quantum dots are located between two metal plates that provide the DC electric field for Stark tuning. We patterned the bottom contact on a silicon carrier chip with electron beam lithography followed by metal deposition (5 nm Cr/50 nm Au) and lift-off. Next, we used a microprobe to transfer an InP waveguide containing InAs quantum dots onto the bottom contact and covered the sample with 500 nm $SiO_2$ deposited by plasma enhanced chemical vapor deposition. Then we formed the top contact by electron beam lithography, followed by 20 nm Cr deposition and lift-off. We chose a thin Cr layer to have optical access through the contact. Finally, we opened a window in the $SiO_2$ by buffer oxide etching to electrically access the buried metal pad connected to the bottom contact. Figure 1b shows a false-color scanning electron microscopy image of the fabricated device.

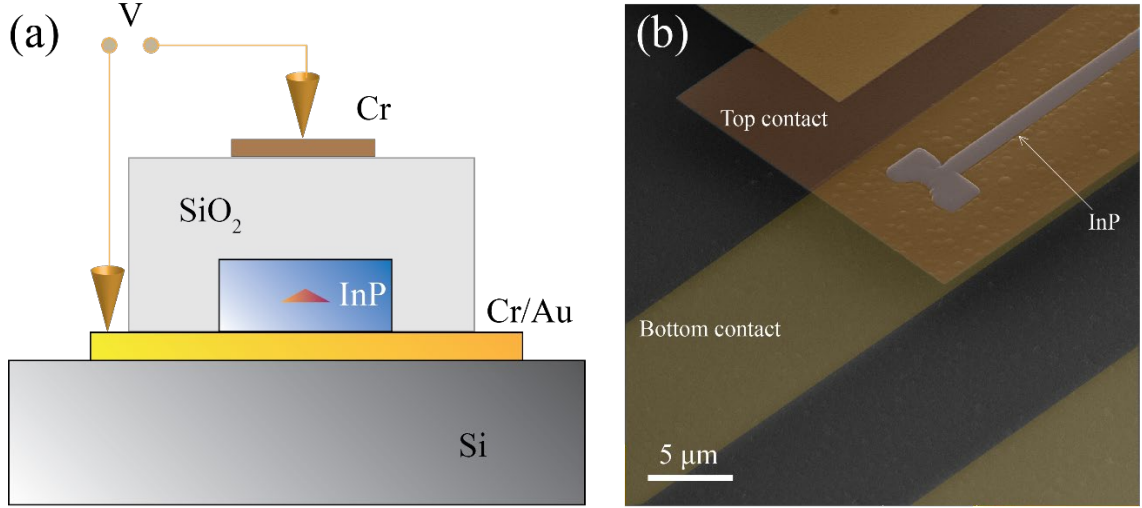

Figure1. (a) An illustration of the proposed Stark tuning device structure. (b) False colored scanning electron microscopy image of the fabricated sample.

We performed photoluminescence measurements on the quantum dots in a closed-cycled cryostat at 4 K. To measure the photoluminescence spectrum of the quantum dots, we used a 780 nm continuous-wave laser for excitation and collected the photoluminescence signal directly through the top metal contact. Figure 2a shows the photoluminescence spectrum of multiple quantum dot emission lines at different DC electric voltages. All quantum dot emission lines start to red shift as we introduce the voltage due to the quantum confined Stark effect.

We extracted photoluminescence emission properties (center wavelength, intensity, and linewidth) from Lorentzian fits to the photoluminescence spectrum at each applied voltage. The relative change in the center wavelength of a quantum dot (QD1) that we labeled in Figure 2a is shown in Figure 2b. We observed a large shift in the emission wavelength of up to 8 nm (5.1 meV) and attribute it to the quantum confined Stark effect. A quadratic fit of the form $\Delta E = pF + \beta F^2$ to this shift allows us to determine the polarizability ($\beta$) and dipole moment ($p$) of the InAs/InP quantum dots. Here $\Delta E$ is the relative change in emission energy and $F$ is the electric field intensity.

From the fit we obtained $\beta = 0.19 \pm 0.01$ $\mu eV \frac{cm^2}{kV^2}$ and $p/e = 0.14 \pm 0.02$ nm, where $e$ is the charge of an electron. These values are slightly lower than the previously reported values for InAs/InP quantum dots.[10,11] Different growth conditions and shape of the quantum dots, parasitic resistance in the device,[19] and non-ideal thickness of the insulator could explain the lower measured values of dipole moment and polarizability compared to the previous works.

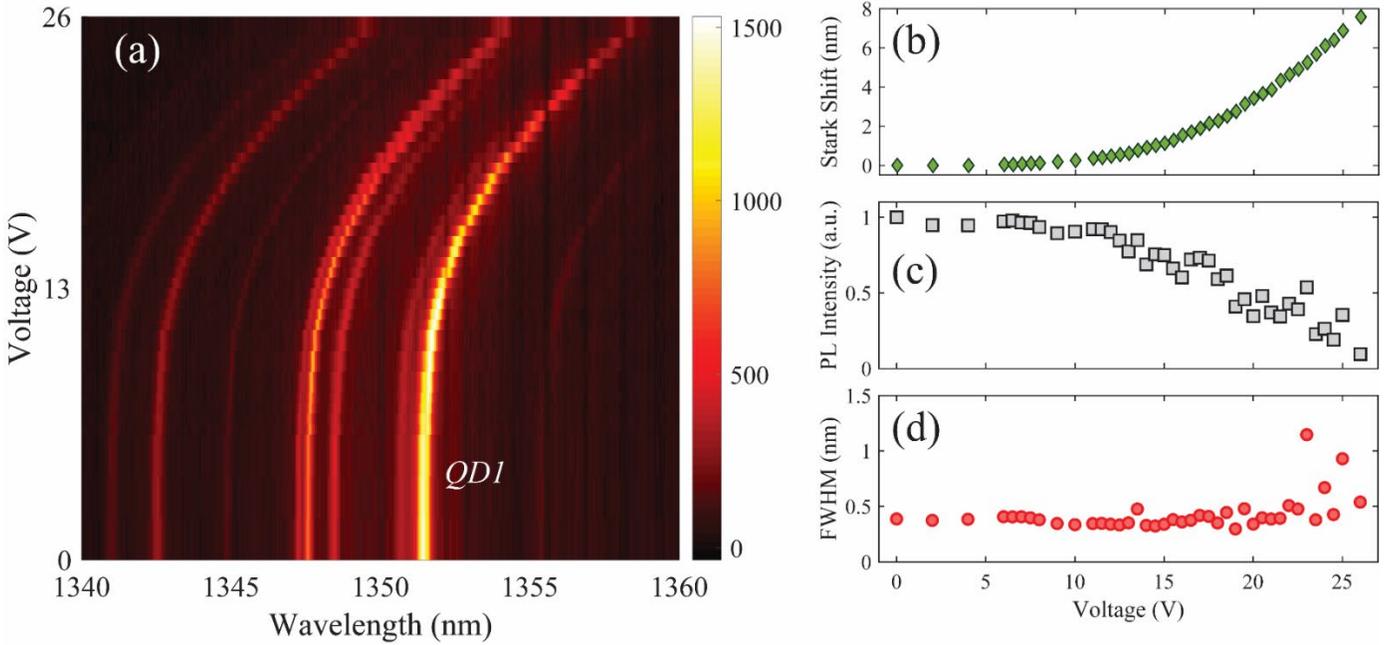

Figure 2. (a) Photoluminescence spectra of several quantum dots for different gate voltages. (b) The Stark shift of QD1. (c) Normalized intensity of QD1. (d) Full width at half maximum for QD1 when fitted to a Lorentzian function.

Figure 2c and 2d show the integrated intensity and linewidth of the QD1 emission as a function of the applied voltage. The integrated intensity of the QD1 emission at 26 V drops to ~10% of its initial value at 0 V applied voltage. At large electric field intensities the created carriers are no longer confined in the quantum dot potential, which leads to tunneling.[20,21] Carrier tunneling reduces the efficiency of emission and therefore explains the drop in integrated intensity. The linewidth of QD1 (Figure 2d) remains constant until very high electric field intensities, indicating the potential of this tuning technique for multi-photon interference applications where narrow linewidths of individual emitters is crucial.

The carrier tunneling effect also represents itself in the excited state lifetime of the quantum dots. Time-resolved photoluminescence measurements reveal changes in the excited state lifetime due to oscillator strength modification as well as tunneling effects.[7] To measure the lifetime of the quantum dot emission, we used a 785 nm pulsed laser with 40 MHz repetition rate. Figure 3a shows the evolution of the excited state lifetime as we changed the applied voltage. Initially, at 0 V the lifetime is 1.44 ns and at higher electric fields (V = 28 V) the lifetime increases to 1.71 ns. We attribute the longer excited state lifetime to the decrease in the oscillator strength of the quantum dot caused by larger separation of the electron-hole pair.[22] At even larger electric field intensities, tunneling effects dominate the excited state lifetime and reduce it to 1.39 ns.[7,22] The rich dynamics of these two competing processes were absent in the photoluminescence intensity measurement (Figure 2c), because both the tunneling effect and oscillator strength reduction lead to a decrease in the emission intensity.

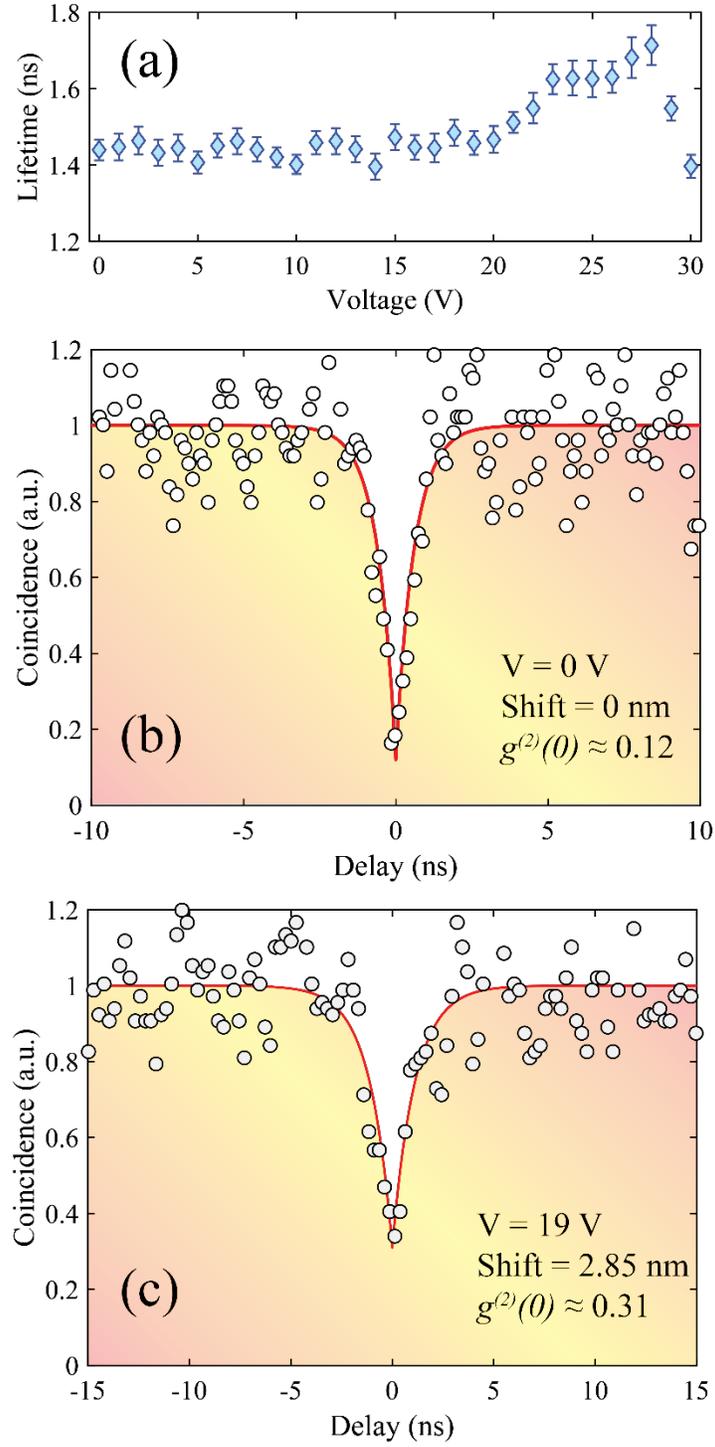

Figure 3. (a) Excited state lifetime of QD1 at different voltages. (b,c) Second-order photon correlation measurement for QD1 at (b) 0 V and (c) 19 V.

Single-photon emission from quantum dots is vital for their application in quantum information processing. Therefore, we performed photon correlation measurements at two different voltages to confirm the single-photon nature of the quantum dot emission and its stability while applying an electric field for Stark tuning. We used a 780 nm continuous-wave laser to excite the quantum dots and sent the photoluminescence signal to a grating spectrometer to filter the emission from QD1. The filtered signal was then sent to a fiber-based beam splitter followed by two superconducting single photon detectors. Figure 3b and Figure 3c show the second-order photon correlation measurements for QD1 at 0 V and 19 V, respectively. Both measurements show clear antibunching behavior, confirming the single photon emission from QD1 even when the emission wavelength was shifted by ~1.8 meV. Exponential fits to the data in Figure 3b and Figure 3c give $g^{(2)}(0) = 0.12$ and $g^{(2)}(0) = 0.31$ for applied voltages of 0 V and 19 V, respectively. The intensity of QD1 at 19 V is about 50% of its value at 0 V (see Figure 2c). To get a second-order correlation measurement at 19 V with similar signal-to-noise ratio as the 0 V measurement, we integrated the histogram 4-times longer. The reduced brightness also degraded the signal to dark count ratio which led to a poorer second order photon correlation measurement of $g^{(2)}(0) = 0.31$. However, this value remains below the classical limit of 0.5, indicating the robustness of single-photon emission against Stark tuning.

In summary, we have demonstrated a large Stark shift in the InAs/InP quantum dot emission wavelength, which is promising for the realization of multiple identical telecom single-photon emitters on a chip. Photon correlation measurements demonstrate robust single-photon emission of the quantum dots before and after Stark tuning. Introduction of wide bandgap barrier layers during the quantum dot growth can reduce the carrier tunneling rates and therefore should further improve the tuning range.[7] Photon collection in this device is based on out-of-plane radiation of the quantum dot through the top metal contact. Using fully transparent top contacts, such as graphene[23] and indium tin oxide,[24] can prevent photon reflection at the top contact and improve the photon collection efficiency. Furthermore, integrating these wavelength-tunable single-photon emitters with integrated photonic waveguides will enable an on-chip optical network with multiple identical quantum emitters for studying multi-photon interference effects,[12] quantum simulation,[25,26] and linear optical quantum computation.[14,15]

ACKNOWLEDGMENT

The authors would like to acknowledge support from the Center for Distributed Quantum Information at the University of Maryland and Army Research Laboratory, and the Physics Frontier Center at the Joint Quantum Institute.